\documentclass{article}

\usepackage{ifthen}

\begin{document}

\title{Analysis Of Measured Transport Properties Of Domain Walls In Magnetic Nanowires And Films}

\author{L. BERGER}

\date{Physics Department, Carnegie Mellon University, Pittsburgh, PA 15213}

\maketitle

\vspace{1ex}

\setlength{\baselineskip}{3.5mm}

Abstract: Existing data for soft magnetic materials of critical current for domain-wall motion, wall speed driven by a magnetic field, and wall electrical resistance, show that all three observable properties are related through a single parameter: the wall mobility $\mu$. The reciprocal of $\mu$ represents the strength of viscous friction between domain wall and conduction-electron gas. And $\mu$ is a function of the wall width, which depends in turn on the aspect ratio t/w, where t and w are the thickness and width of the sample. Over four orders of magnitude of $\mu$, the data for nanowires show $\mu\propto (t/w)^{-2.2}$. This dependence is in approximate agreement with the prediction of the 1984 Berger theory based on s-d exchange. On the other hand, it is inconsistent with the prediction of the 2004 Tatara and Kohno theory, and of the 2004 Zhang and Li theory.

\vspace{2ex}


\vspace{2ex}

I. INTRODUCTION

\vspace{1ex}

Many years ago, we predicted $^{1}$ that a dc current would apply a force on a domain wall, through the s-d exchange interaction. We also observed $^{2,3}$ wall displacements caused by such a force, in Ni-Fe films containing Neel walls. As predicted, the wall displacements had the same sense as the drift speed of the electron-like carriers. Gan et al. $^{4}$ saw similar current-induced wall displacements in Ni-Fe films with smaller patterned width of 20 $\mu m$. Since then, the same kind of results have been reported by many authors for magnetic nanowires $^{5-14}$. These nanowires, mostly made of Ni-Fe, have a much larger ratio of thickness t to width w, resulting in thinner walls and, consequently $^{1}$, to a stronger interaction of wall and electron gas. The values of t and w, of the measured critical current density $j_{c}$ where wall motion starts in zero field, and of the measured coercive field $H_{c}$ for wall motion at zero current are the important measured quantities. In some cases, $j_{c}$ or $H_{c}$ have to be inferred from more indirect information given in the paper.

Ebels et al. $^{15}$ as well as Khizroev et al. $^{16}$ and Lepedatu and Xu $^{5}$ have measured the extra electrical resistance $R_{dw}$ caused by a wall, in nanowires. The Ebels cobalt sample had a circular cross section with t=w as the diameter. 

The wall speed $v_{w}$ induced by an easy-axis field H is proportional to H, and the ratio $\mu = v_{w}/H$ is called the wall mobility. Patton and Humphrey $^{17}$ as well as Shishkov $^{18}$ have measured $\mu$ for walls in Ni-Fe films over a range of thickness t. They also summarized earlier work. We use their results for $t<1\mu m$, where eddy-current damping is negligible. Freitas and Berger $^{2}$ have measured $\mu$ in Ni-Fe in the same samples where they moved walls with a current. Finally, Ono et al.$^{19}$ measured $\mu$ on a Fe-Ni nanowire.

The purpose of the present paper is to show that the concept of wall mobility can be used to characterize in a consistent manner the results of experiments of current-induced wall motion, or of measurements of wall resistance, just as well as those of field-induced wall motion. No matter the origin of a value of $\mu$, we find that value to be the same function of the ratio t/w of thickness to width of the sample. The physical reason $^{1}$ is that in all three cases we have viscous friction between a domain wall and the electron gas. The value of the friction force between the two is the same whether the electron gas is moving with respect to an initially static wall (experiments of wall resistance or of current-induced wall motion), or whether the wall moves and the gas is at rest (field-induced wall motion).

We also show that these experiments agree with the 1984 Berger theory, but not with the 2004 theories by Tatara and Kohno or by Zhang and Li.

\vspace{1ex}

II. WALL MOBILITY DERIVED FROM 

CRITICAL CURRENT OR WALL RESISTANCE 

\vspace{1em}

We start with the force $F_{H}$ exerted by an easy-axis field H on a $180^{\circ}$ wall, given $^{20}$ by
\begin{equation}
 F_{H}=2\mu_{0}M_{s}Htw.
\end{equation}

 In the steady state, this force must be balanced by a viscous damping force $F_{d}$, giving $F_{H}+F_{d}=0$. Combining this equation and Eq. (1) with the mobility definition $\mu=v_{w}/H$ given earlier, we obtain

\begin{equation}
F_{d}=-2\mu_{0}M_{s}tw\mu^{-1}v_{w}.
\end{equation}

In the 1984 Berger theory $^{1}$, the origin of this damping force is the friction of the moving wall with the electron gas at rest. If, instead, the gas is moving because an electric current density j is present, the average drift speed of the gas is $v_{e}=-j/n_{e}e=jR_{0}$, where $n_{e}$ is the electron density in a one-band model, and $R_{0}$ the ordinary Hall coefficient. If the interaction depends only on relative motion, we expect $^{1}$ $v_{e}$ to generate a drive force $F_{e}$ on the wall similar to the damping force $F_{d}$ generated by $v_{w}$ (see Eq. (2)), though of the opposite sign:

\begin{equation}
F_{e}=2\mu_{0}M_{s}tw\mu^{-1}\beta_{1}v_{e}.
\end{equation}

Here, $\beta_{1}$ is $^{1}$ a correction factor found experimentally $^{2,3}$ to be equal to about 2 for Ni-Fe films.

There is also a pinning force $F_{p}$ applied to the wall by lattice defects. By analogy with Eq. (1), it is

\begin{equation}
F_{p}=\pm 2\mu_{0}M_{s}H_{c}tw.
\end{equation}

where $H_{c}$ is the coercive field. At the critical current density $j_{c}$ for wall motion, $F_{e}$ just balances $F_{p}$, with $F_{e}+F_{p}=0$. Combining this equation and the relation $v_{e}=R_{0}j$ with Eqs. (3) and (4), we obtain finally:

\begin{equation}
j_{c}=\frac{\mu H_{c}}{\beta_{1}|R_{0}|}.
\end{equation}

From the measured $j_{c},H_{c}$ obtained from Refs. 2-14, we find $\mu$ for films and nanowires by Eq. (5), assuming $^{2}$ $R_{0}=-1.36\times10^{-10}m^{3}/C$ for $Ni_{80}Fe_{20}$ and $^{21}$ $-1.1\times10^{-10}m^{3}/C$ for Co, also $^{21}$ $-1.25\times 10^{-10}m^{3}/C$ for $Fe_{50}Co_{50}$. We plot $\mu$ versus t/w in Fig. 1 for all these samples. In several cases, the quantities actually measured were not $j_{c},H_{c}$ but critical-current variations $\Delta j_{c}$ corresponding to field variations $\Delta H$. But Eq. (5) keeps the same form.

To the drive force $F_{e}$ (Eq. (3)) exerted by electrons on the wall corresponds a reaction force $-F_{e}$ applied to the electron gas, which manifests itself $^{22}$ as a wall resistance $R_{dw}>0$:

\begin{equation}
R_{dw}=\frac{2\beta_{1}\mu_{0}M_{s}R^{2}_{0}}{\mu wt}.
\end{equation}

From the measured $R_{dw}$ for nanowires $^{5, 15, 16}$, we find $\mu$ by Eq. (6), and plot it versus t/w in Fig. 1. We assume $\mu_{0}M_{s}=1T$ for $Ni_{80}Fe_{20}$ and 1.8 T for Co, and the same $\beta_{1},R_{0}$ as before. Note that Eq. (6) does not hold in the case of $R_{dw}$ mechanisms where the momentum flows from the electrons to the lattice, rather than to the wall itself. One example is the mechanism of Gregg and of Levy and Zhang $^{23}$. Another is the anisotropic magnetoresistance. Because of this, we have only used $R_{dw}$ measurements where the $R_{dw}$ value predicted by Eq. (6) is larger than the Gregg or anisotropic-magnetoresistance ones, i.e., for small $\Delta_{0}$ corresponding to large $t/w>0.1$. Indeed, the Gregg resistance $^{23}$ varies like $\Delta_{0}^{-1}\propto (t/w)^{1/2}$, slower than our $R_{dw}$ of Eq. (6).

\vspace{1ex}

III. EMPIRICAL RELATION BETWEEN

 WALL MOBILITY AND ASPECT RATIO

\vspace{1ex}

The wall mobilities $\mu$ obtained by the three different methods are plotted in Fig. 1 versus t/w, for all films and nanowires, on double logarithmic scales. The fact that the directly measured wall mobilities (one case for nanowires and many for films) are roughly consistent with those derived from critical current (many for nanowires and three for films) and with those from wall resistance (six for nanowires) indicates that the phenomenological idea $^{1}$ of viscous friction between electron gas and domain wall, used in the preceding data analysis, is sound.

 We see (Fig. 1) that the $\mu$ values for nanowires start at the same level as those for films, but plunge with a large negative slope, dropping by four orders of magnitude as t/w increases. Although the dispersion of the data points is rather large, the best fit of a straight line to the nanowire data is with a slope of -2.2 (dashed line in Fig. 1). Only the data of Klaui et al. $^{14}$ really differ from this scheme. The average horizontal deviation of data points from the dashed line is 2.8 decibels, excluding the point of Ref. 14. This compares to a total range of variation of t/w by a factor of 100, or 20 decibels, leading to an accuracy of about 14\% on the slope of the dashed line. The correlation coefficient of the nanowire data is found to be 80\%.

 The film data (Fig. 1), including those of Gan et al. (Ref. 4), seem to lie in a nearly horizontal band. The reason is probably that t/w is too small for the shape anisotropy to dominate over the induced anisotropy, so that the wall width varies little with t/w, except near $t/w=10^{-5}$ where the walls change from Neel to Bloch type.

We still need to find the relation between $\Delta_{0}$ and t/w in the case of nanowires, where the main contribution to the anisotropy energy is $^{25}$ the shape anisotropy. The latter is connected with magnetic poles at the wire surface. For the simplest kind of head-to-head $180^{\circ}$ domain wall with no vortices, we obtain $\Delta_{0}$ by minimizing the sum of exchange and anisotropy energies:

\begin{equation}
\Delta_{0}=\frac{l_{ex}}{(\frac{2}{\pi}arctan(\frac{t}{w}))^{1/2}}.
\end{equation}

Here, $l_{ex}=(2A/\mu_{0}M_{s}^{2})\simeq 6nm$ is the exchange length of $Ni_{80}Fe_{20}$, and $A\simeq 1.5\times10^{-11}J/m$ the exchange stiffness. The wall mobility is often written $^{24}$ in terms of the Gilbert damping parameter $\alpha$ as $\mu=\gamma\Delta_{0}/\alpha$, where $\gamma$ is the gyromagnetic factor. The results of Fig. 1 for nanowires imply that, like $\mu$, $\alpha$ is dependent on t/w through the wall width $\Delta_{0}$, with $\alpha\propto(t/w)^{1.7}\propto \Delta_{0}^{-3.4}$ if $t/w\ll 1$.

\vspace{1ex}

IV. MICROSCOPIC THEORY OF VISCOUS FORCES

\vspace{1ex}

We now turn to the 1984 microscopic theory $^{1}$, which was based on the force applied to conduction-electron spins by the gradient of the s-d exchange field present in the wall. This force between electrons and wall is similar to the one involved in the famous Stern-Gerlach experiment, but with the magnetic field replaced by an exchange field. It is important not to confuse this force (momentum transfer) with the spin-transfer torque, which is not equivalent to a force and was discussed in the 1986 paper of Ref. 22. The reciprocal of the wall mobility is predicted to be:

\begin{equation}
\mu^{-1}=\frac{\tau_{\uparrow\downarrow}D_{4s}}{16\mu_{0}M_{s}}\int^{+\infty}_{-\infty}dx[\frac{\partial}{\partial x}(V_{\uparrow}(x)-V_{\downarrow}(x)]^{2}.
\end{equation}

Here, $\tau_{\uparrow \downarrow}$ is $^{1}$ the relaxation time of electrons between bands. It is largest in the particular case of spin-up and spin-down bands, considered here, where it becomes the spin-relaxation time. One of the two $\partial(V_{\uparrow}(x)-V_{\downarrow}(x))/\partial x$ factors, when combined with the $\tau_{\uparrow\downarrow}$ factor, represents a non-equilibrium transfer of electron density between the two bands, i.e., a spin accumulation. This spin accumulation is a scalar which has $^{1}$ opposite signs on the two sides of the domain wall. Also, $D_{4s}$ is the density of states of 4s conduction electrons, assumed the same for spin-up and spin-down states. And $V_{\uparrow}(x)>0, V_{\downarrow}(x)<0$ are potential-energy functions $^{1}$ of maximum order $10^{-3}eV$ which vanish outside the wall. If spin-up and spin-down conduction electrons have the same Fermi velocity $v_{F}$, we have

\begin{equation}
V_{\uparrow}(x)=-V_{\downarrow}(x)=J_{sd}S(1-cos\gamma(x));
\gamma(x)=arctan(\frac{\hbar v_{F}}{2SJ_{sd}3^{1/2}}\frac{d\theta}{dx}).
\end{equation}

Here, $J_{sd}$ is the s-d exchange integral, and S the atomic spin of magnetic 3d electrons assumed localized. Also, $\theta(x)$ is the angle of the local magnetization in the wall with the domain magnetization, given $^{20}$ by $x/\Delta_{0}=ln\ tan(\theta/2)$. Finally, $\gamma(x)$ is the local angle between conduction-electron spin and magnetic spin in the wall, later called ``mistracking angle'' by Gregg $^{23}$. Semiclassical equations of motion $^{1}$ for conduction-electron spins are used to derive Eqs. (9). While the approximation $\gamma(x)\ll 1rad.$ was made in Ref 1, Eqs. (9) are valid even when $\gamma(x)\simeq 1rad.$. Eqs. (9) show that $V_{\uparrow}$ and $V_{\downarrow}$ arise from the existence of the mistracking angle.

It was shown in Ref. 1 that the Stern-Gerlach forces on individual electrons of spin up and spin down can be written in the form of $-\partial V_{\uparrow}/\partial x,-\partial V_{\downarrow}/\partial x$, respectively. This leads to the second $\partial (V_{\uparrow}(x)-V_{\downarrow}(x))/\partial x$ factor in Eq. (8). The spin accumulation mentioned above is needed if the total force created by the left half of the wall is not to cancel the total force of the right half (see Fig. 1 of Ref. 1). 

We evaluate $\mu$ numerically as a function of t/w by combining Eqs. (7-9). We use values $J_{sd}=0.6eV$, and $\mu_{0}M_{s}=1T, S=0.5$ for $Ni_{80}Fe_{20}$. We derive the spin-relaxation time $\tau_{\uparrow\downarrow}$ from the spin-diffusion length $l_{sr}=(2\tau_{\uparrow\downarrow}/e^{2}D(\rho_{\uparrow}+\rho{\downarrow}))^{1/2}$ where D is the full density of states including 3d electrons, and $\rho_{\uparrow},\rho_{\downarrow}$ the spin-up, spin-down resistivities. The electronic specific heat $^{26}$ of $Ni_{80}Fe_{20}$ gives $D=1.09\times10^{48}J^{-1}m^{-3}$. We assume $\rho_{\uparrow}+\rho_{\downarrow}=133\times10^{-8}\Omega m$ for $Ni_{80}Fe_{20}$ at room temperature, somewhat higher than the $105\times10^{-8}\Omega m$ derived from the measured low-temperature $\rho_{\uparrow},\rho_{\downarrow}$ values $^{27}$. Finally, using a value $l_{sr}=60 nm$ for a ferromagnet such as cobalt $^{28}$, we obtain $\tau_{\uparrow\downarrow}=6.7\times10^{-11}s$.

If we use free-electron values $v_{F}=1.57\times10^{6}m/s,D_{4s}=11.4\times10^{46}J^{-1}m^{-3}$, suitable $^{29}$ for conduction electrons in Cu, then Eqs. (7-9) yield much too large $\mu$ values for given t/w, corresponding to very small viscous coupling between electrons and wall. In order to fit the nanowire data of Fig. 1, we are forced to use $v_{F}= 22.2\times 10^{6}m/s$, about 14 times larger than the free-electron value above. To be consistent, we have to reduce $D_{4s}$ by the same factor, to $8.1\times 10^{45}J^{-1}m^{-3}$. The $\mu$ predicted by Eqs.(7-9) are then plotted versus t/w as the solid curve in Fig. 1. The too large needed $v_{F}$ value mentioned above may represent a genuine band-structure effect in nanowires. But if, instead of $l_{sr}=60nm$, we used the surprisingly short $\simeq 4-5nm$ measured $^{27}$ in Ni-Fe films, an even larger $v_{F}$ value would be needed.

We see (Fig. 1) that the fit between theory (solid curve) and the average of experiments (dashed line) is rather good. In the limit $\gamma(x)\ll 1$, realized at small t/w, Eq. (9) gives $\gamma(x)\propto \Delta_{0}^{-1}$ and $V_{\uparrow},V_{\downarrow}\propto \Delta_{0}^{-2}$. Plugging this into Eq. (8), we obtain $\mu\propto\Delta_{0}^{5}$. Using Eq. (7), this means $\mu\propto (t/w)^{-2.5}$, giving a predicted slope is -2.5. As t/w and $\gamma(x)$ increase, the slope of the theoretical curve becomes less negative and is close, overall, to the average experimental slope of -2.2 indicated by the dashed line (Fig. 1). The mistracking angle $\gamma(x)$ changes from 0.19 rad. to 1.0 rad. as t/w increases over the range covered by the nanowires. Note that $\mu \propto (t/w)^{-2.2}$ implies $\mu \propto \Delta_{0}^{4.4}$ by Eq. (7), if $t/w\ll 1$.  

\vspace{1ex}

V. IMPORTANCE OF RANDOM SCATTERING

\vspace{1ex}

In $Ni_{80}Fe_{20}$ films or nanowires, the average of spin-up and spin-down electron mean free paths has been estimated $^{27}$ at $\simeq 2nm$. In Co, it is not much longer. This justifies the 1984 Berger theory $^{1}$, where the mean free path is assumed smaller than the wall width $\Delta_{0}=8-70nm$, given by Eq. (7) for the present nanowires. The small mean free path breaks up the coherence of the wavefunction, leading $^{1}$ to the validity of a semiclassical transport theory with a local conductivity defined at each point inside the wall

Tatara and Kohno's recent theory $^{30}$, like the Berger theory $^{1}$, is based on the ``Stern-Gerlach force'' caused by the gradient of the exchange field in the wall. However, they do not have any random electron scattering by solutes or phonons, so that their wavefunction is highly coherent over the whole wall. This leads to nearly-complete destructive interference between wavelets reflected at different points of the wall, resulting in a very small total force which varies exponentially $^{30}$ with wall width $\Delta_{0}$, not as a power law as for the data of Fig. 1 or for  the 1984 Berger theory. This is shown as the dotted curve labelled T-K in Fig. 1, assuming $J_{sd}=0.1eV,v_{F}=1.57\times 10^{6}m/s$.
   
 The theory by Thiaville et al. $^{31}$ is based on the same phenomenological idea as our data analysis (Eqs. (2-3)), namely that the electron drift speed $v_{e}$ generates friction forces of the same magnitude as the wall speed $v_{w}$.

 Zhang and Li $^{32}$ predict a force proportional to $\Delta_{0}^{-1}$. This is equivalent to a wall mobility $\mu\propto \Delta_{0}\propto(t/w)^{-0.5}$, shown as a dotted line of slope -0.5 in Fig. 1, labelled Z-L. This is clearly inconsistent with the $\mu\propto(t/w)^{-2.2}$ found experimentally by us in nanowires (dashed line in Fig. 1), and predicted by the Berger theory (solid curve in Fig. 1). In addition, the $\mu$ values predicted by Zhang and Li for values  $\tau_{\uparrow\downarrow}=0.7\times 10^{-13}s$ and $J_{sd}=0.6eV$ (Fig. 1) are too large by at least three orders of magnitude to explain the experimental results for $t/w > 0.1$. They correspond to a very small force on the wall. As in the case of the Tatara and Kohno theory, we chose the above parameter values in such a way as to maximize the force.

Very recently, we learnt about a calculation by Zhao et al. $^{33}$ of the resistance of a domain wall at a point contact in the diffusive limit. Using our Eq. (6), their result is equivalent to a mobility $\mu \propto \Delta_{0}^{2} \propto (t/w)^{-1}$.
\vspace{1ex}

VI. CONCLUSIONS

\vspace{1ex}

We found that measured domain-wall electrical resistance and current-indu\-ced as well as field-induced wall motion are all related through the concept of wall mobility $\mu$, and lead to the same dependence $\mu \propto(t/w)^{-2.2}$ on the aspect ratio t/w of the cross section of a magnetic nanowire (Fig. 1). Note that, for domain-wall resistance, this applies only at $t/w>0.1$, where the resistance mechanism of Eq. (6) dominates. This dependence of $\mu$ on t/w agrees with the prediction (Eqs. (7-9)) of the 1984 Berger theory $^{1}$, though a too large value of the Fermi velocity is needed to explain the observed $\mu$ values. But it is inconsistent with the 2004 Tatara and Kohno theory, as well as with the 2004 Zhang and Li theory.

I am grateful to Sonali Mukherjee for useful discussions on this topic.

\vspace{1ex}

REFERENCES

\vspace{1ex}

1. L. Berger, J. Appl. Phys. \textbf{55}, 1954 (1984).

2. P.P. Freitas and L. Berger, J. Appl. Phys. \textbf{57}, 1266 (1985).

3. C.Y. Hung and L. Berger, J. Appl. Phys. \textbf{63}, 4276 (1988).

4. L. Gan, S.H. Chung, K.H. Aschenbach, M. Dreyer, and R.D. Gomez, I.E.E.E. Trans. Magn. \textbf{36}, 3047 (2000).

5. S. Lepadatu and Y.B. Xu, Phys. Rev. Lett. \textbf{92}, 127201 (2004); S. Lepedatu, Y.B. Xu, and E. Ahmad, J. Appl. Phys. \textbf{97}, 10C711 (2005).

6. J.L. Tsai, S.F. Lee, Y. Liou, Y.D. Yao, T.Y. Chen, and K.W. Cheng, J. Appl. Phys. \textbf{97}, 10C710 (2005).

7. T. Kimura, Y. Otani, K. Tsukagoshi and Y. Aoyagi, J. Appl. Phys. \textbf{94}, 7947 (2003).

8. T. Kimura, Y. Otani, I. Yagi, K. Tsukagoshi and Y. Aoyagi, J. Appl. Phys. \textbf{94}, 7266 (2003).

9. C.K. Lim, T. Devolder, C. Chappert, J. Grollier, V. Cros, A. Vaures, A. Fert, G. Faini, Appl. Phys. Lett. \textbf{84}, 2820 (2004).

10. J. Grollier, P. Boulenc, V. Cros, A. Hamzic, A. Vaures, and A. Fert, Appl. Phys. Lett. \textbf{83}, 509 (2003).

11. A. Yamaguchi, S. Nosu, H. Tanigawa, T. Ono, K. Miyake, K. Mibu, and T. Shinjo, Appl. Phys. Lett. \textbf{86}, 012511 (2005); A. Yamaguchi, T. Ono, S. Nasu, K. Miyake, K. Mibu, T. Shinjo, Phys. Rev. Lett. \textbf{92}, 077205 (2004).

12. M. Tsoi, R.E. Fontana and S.S.P. Parkin, Appl. Phys. Lett. \textbf{83}, 2617 (2003).

13. N. Vernier, D.A. Allwood, D. Atkinson, M.D. Cooke, and R.P. Cowburn, Europhys. Lett. \textbf{65}, 526 (2004).

14. M. Klaui, C.A.F. Vaz, J.A.C. Bland, W. Wernsdorfer, G.Faini, E. Cambril, and L.J. Heyderman, Appl. Phys. Lett. \textbf{83}, 105 (2003).

15. U. Ebels, A. Radulescu, Y. Henry, L. Piraux and K. Ounadjela, Phys. Rev. Lett. \textbf{84}, 983 (2000).

16. S. Khizroev, Y. Hijazi, R. Chomko, S. Mukherjee, R. Chantrell, X. Wu, R. Carley, and D. Litvinov, Appl. Phys. Lett. \textbf{86}, 042502 (2005).

17. C.E. Patton and F.B. Humphrey, J. Appl. Phys. \textbf{37}, 4269 (1966).

18. A.G. Shishkov, Czechosl. J. Phys. \textbf{B21}, 368 (1971).

19. T. Ono, H. Miyajima, K. Shigeto, K. Mibu, N. Hosoito, and T. Shinjo, Science \textbf{284}, 468 (1999).

20. S. Chikazumi, \textbf{Physics of Magnetism} (Wiley, N.Y., 1964), p. 272.

21. N.V. Volkhenstein and G.V. Fedorov, Sov. Phys.-JETP \textbf{11}, 48 (1960); S. Foner and E.M. Pugh, Phys. Rev. \textbf{91}, 20 (1953); F.P. Beitel and E.M. Pugh, ibido \textbf{112}, 1516 (1958); S. Foner, F.E. Allison and E.M. Pugh, ibido \textbf{109}, 1129 (1958).

22. L. Berger, J. Phys. Chem. Solids \textbf{35}, 947 (1974). See Eq. (28); L. Berger, Phys. Rev. B \textbf{33}, 1572 (1986). See Eq. (15).

23. J.F. Gregg, W. Allen, K. Ounadjela, M. Viret, M. Hehn, S.M. Thompson and J.M.D. Coey, Phys. Rev. Lett. \textbf{77}, 1580 (1996); P.M. Levy and S. Zhang, ibido \textbf{79}, 5110 (1997).

24. A.P. Malozemoff and J.C. Slonczewski, \textbf{Magnetic Domain Walls In Bubble Materials}, (Academic Press, N.Y., 1979). See p. 128.

25. H.B. Braun, J. Appl. Phys. \textbf{85}, 6172 (1999); S. Mukherjee, D. Litvinov and S. Khizroev, IEEE Trans. Magn. \textbf{40}, 2143 (2004).

26. L. Berger, Physica \textbf{91B}, 31 (1977). See Fig. 2.

27. S. Dubois, L. Piraux, J.M. George, K. Ounadjela, J.L. Duvail, and A. Fert, Phys. Rev. B \textbf{60}, 477 (1999); S.D. Steenwyk, S.Y. Hsu, R. Loloee, J. Bass, and W.P. Pratt, J. Magn. Magn. Mater. \textbf{170}, L1 (1997).

28. L. Piraux, S. Dubois, A. Fert, and L. Beliard, Eur. Phys. J. B \textbf{4}, 413 (1998).

29. C. Kittel, \textbf{Introduction To Solid State Physics}, (Wiley, N.Y., 1986), 6th edition, p. 134, 141.

30. G. Tatara and H. Kohno, Phys. Rev. Lett. \textbf{92}, 086601 (2004).

31. A. Thiaville, Y. Nakatani, J. Miltat and Y. Suzuki, preprint cond-mat/0407628 v2.

32. S. Zhang and Z. Li, Phys. Rev. Lett. \textbf{93}, 127204 (2004).

33. Y.W. Zhao, M. Munoz, G. Tatara, and N. Garcia, J. Magn. Magn. Mater. \textbf{223}, 169 (2001).

\vspace{1ex}

        FIGURE CAPTIONS

\vspace{1em}

FIG. 1: Wall mobility $\mu$ versus aspect ratio t/w of cross section of sample. Plain circles represent $\mu$ values from measurements of critical current density $j_{c}$. Circles with a tail to the left represent $\mu$ values from measurements of domain-wall resistance $R_{dw}$, and those with a tail to the right direct measurements of $\mu$. The numbers in the circles indicate the references from which the data come. The dashed straight line is fitted to the nanowire data, and has a slope of -2.2. The solid curve is the prediction of the 1984 Berger theory (see Eqs. (7-9)). The two dotted curves are the predictions of the Tatara and Kohno theory and of the Zhang and Li theory.
 
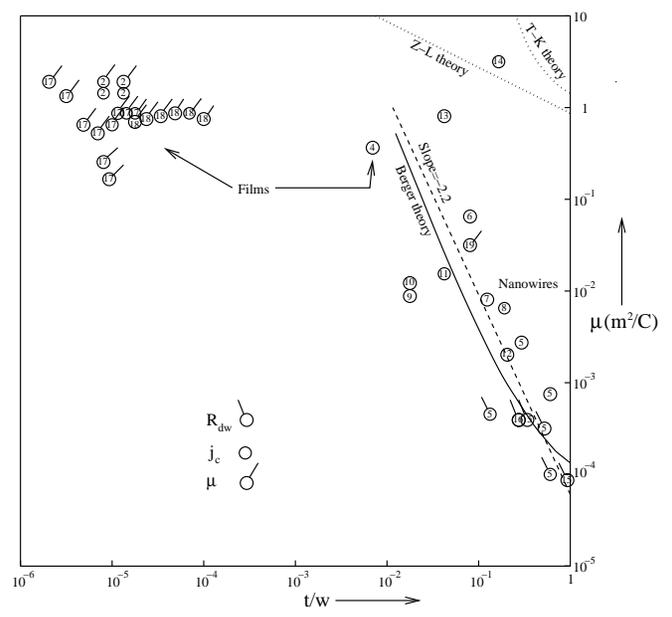
\end{document}